\documentstyle[aps,epsf,prl,multicol]{revtex}

\begin{document}
\draft
\title{Localization and Mobility Edge in One-Dimensional Potentials\\
with Correlated Disorder}
\author{
F.M.~Izrailev  and A.A.~Krokhin}

\address{
Instituto de F\'isica, Universidad Autonoma de Puebla, Apdo. Postal J-48,
Puebla, 72570 M\'exico
}

\date{\today}
\maketitle

\begin{abstract}
We show that a mobility edge exists in 1D random potentials provided
specific long-range
correlations. Our approach is based on the relation between binary correlator
of a site potential and the localization length.
We give the algorithm to construct numerically potentials with mobility
edge at any given energy inside allowed zone. 
Another natural way to generate such potentials
is to use chaotic trajectories of non-linear maps.
Our numerical calculations for few particular potentials demonstrate
the presence of mobility edges in 1D geometry.
\end{abstract}

\pacs{PACS numbers: 72.15.Rw, 03.65.BZ, 72.10.Bg}

\begin{multicols}{2}

It is commonly believed that there is no mobility edge in 1D models with
random-like potentials. This is based on the fact that for random potentials
all eigenstates are exponentially localized, no matter how weak the
randomness is \cite{A58}. On the other hand, for potentials with
``correlated disorder'' the localization length diverges for specific values
of energy (see, e.g. \cite{F89,B92}). The
well-studied model of this kind is the so-called ``random dimer'' \cite
{DWP90} for which the potential has peculiar short-range correlations.
Though there is no mobility edge for such potentials, this example shows a
highly non-trivial role of correlations. In this Letter we study the
relation between correlations in the site-potential of 1D tight-binding
model and localization properties of eigenstates, and give examples of the
potentials with mobility edges inside the energy band.

The model under consideration is the discrete Shr\"odinger equation for
stationary eigenstates $\psi_n (E)$ ,

\begin{equation}
\label{base}\psi _{n+1}+\psi _{n-1}=(E+\epsilon _n)\,\psi _n\,\,,
\end{equation}
where $E$ is the eigenenergy and $\epsilon _n\,$ is the site-potential. 
To study the origin of delocalized states in long-correlated random 
potentials, 
we suggest a simple and clear approach based on the 
representation of the quantum model (\ref{base}) in terms of 
classical two-dimensional Hamiltonian map,
\begin{equation}
\label{map1}
\begin{array}{c}
p_{n+1}=p_n+(E-2-\epsilon _n)\,x_n, \\ 
x_{n+1}=x_n+p_{n+1},
\end{array}
\end{equation}
where $p_{n+1}=x_{n+1}-x_n\,$and $\,x_n=\psi _n$. This map describes the
behavior of a linear oscillator subjected to linear periodic delta-kicks
with amplitude depending on $\epsilon _n$. In such an approach, localized
quantum states correspond to trajectories which are unbounded in the
classical phase space $(p,x)$ when $n\rightarrow \infty $. Contrary,
extended states are represented by bounded trajectories.

It is convenient to introduce the action-angle variables $(r,\theta )$
and represent Eqs.(\ref{base}) and (\ref{map1}) in the followind form
(see details in \cite{IKT95}),
\begin{equation}
\label{mapbase}
\begin{array}{c}
\sin \,\theta _{n+1}=D_n^{-1}\left( \sin \,(\theta _n-\mu )\,-\,A_n\sin
\,\theta _n\sin \,\mu \right),  \\ 
\cos \,\theta _{n+1}=D_n^{-1}\left( \cos \,(\theta _n-\mu )\,+\,A_n\sin
\,\theta _n\cos \,\mu \right), 
\end{array}
\end{equation}
where 
\begin{equation}
\label{Dn}
\begin{array}{c}
D_n=
\frac{r_{n+1}}{r_n}=\sqrt{1+A_n\sin \,({2\theta _n)}+A_n^2\sin {}^2\theta _n}%
\;, \\  \\ 
\quad A_n=-\epsilon _n/\sin \mu \;,\quad E=2\cos \mu \,.
\end{array}
\end{equation}
We use the following definition of the inverse localization length (or
Lyapunov exponent $\Lambda $) \cite{IKT95}, 
\begin{equation}
\label{locdef}l^{-1}\equiv \Lambda =\lim _{N\rightarrow \infty }\,\frac
1N\sum_{n=1}^N\ln \left( \frac{r_{n+1}}{r_n}\right) \,\,,
\end{equation}
which coincides inside the energy band, $|E|<2$
with the standard definition \cite{LGP88}, $\Lambda =\langle
\ln |\frac{{\psi _{n+1}}}{{\psi_n }}|\rangle $ 
(see details in \cite{IRT98}). Here the brackets stand for the average over $%
n$. This Hamiltonian map approach turns out to be very effective in the
study of completely disordered potentials \cite{IRT98} as well as potentials
with correlated disorder \cite{IKT95,KTI97}.

Below we consider a general case of {\it any} stationary site-potential $%
\epsilon _n$ under the condition $\left| {\epsilon _n}\right| <<1$. In this
case one can expand the logarithm in Eq. (\ref{locdef}) and in the second
order of perturbation theory gets, 
\begin{equation}
\label{locmain}l^{-1}={\frac{{\left\langle {\epsilon _n^2}\right\rangle }}{%
8\,{\sin {}^2\mu }}}-{\frac{{\left\langle {\epsilon _n\sin \,({2\theta _n)}}%
\right\rangle }}{2\,{\sin \mu }}.} 
\end{equation}

In order to calculate the correlator $\left\langle \epsilon {_n\sin \,({%
2\theta _n)}}\right\rangle $ with quadratic accuracy, we use the approximate
one-dimensional map for the phase $\theta _n$ obtained from Eq. (\ref
{mapbase}),

\begin{equation}
\label{tetaN}\theta _n=\theta _{n-1}-\mu +\epsilon {_{n-1}\frac{\sin
{}^2\theta _{n-1}}{{\sin \mu }}}\;. 
\end{equation}
From Eq.~(\ref{tetaN}) the correlator $\left\langle \epsilon {_n\sin
\,({2\theta _n)}}\right\rangle $ can be expressed through the precedent one, 
$\left\langle {\epsilon _n\sin \,2({\theta _{n-1})}}\right\rangle $. 
Let us introduce the following notations, 
\begin{equation}
\label{corr} 
\begin{array}{c}
Q_k\equiv \left\langle 
{\epsilon _ne^{2i\theta _{n-k}}}\right\rangle =-\frac i2\frac{\exp (-2i\mu ) 
}{\sin \mu }\,\epsilon _0^2a_k\;, \\  \\ 
\quad \left\langle \epsilon {_n\epsilon _{n-k}}\right\rangle =\epsilon
_0^2\xi (k)\;,\quad \left\langle \epsilon {_n^2}\right\rangle =\epsilon
_0^2\;. 
\end{array}
\end{equation}
It can be shown that the normalized correlators $a_k$ are defined by the
infinite set of linear equations, 
\begin{equation}
\label{matrix}a_{k-1}\,-e^{-2i\mu }\,a_k=\xi (k),\,\,\,1\leq k<\infty . 
\end{equation}
These equations emerge after multiple application of the recursion relation (%
\ref{tetaN}) to the correlator $Q_k$.

According to Eq. (\ref{locmain}) the localization length is determined by
the only term $a_0$, 
\begin{equation}
\label{locRe}l^{-1}={\frac{\epsilon _0^2}{8\,{\sin {}^2\mu }}}+ {\frac
{\epsilon _0^2} {4 \,{\sin {}^2\mu }}}Re\,\left( e{^{-2i\mu }}a_0\right) \,
\, , 
\end{equation}
which can be easily obtained from Eq. (\ref{matrix}), 
\begin{equation}
\label{a0}a_0=\sum\limits_{k=1}^\infty \xi (k)\,\exp \left[ -2i\mu
(k-1)\right]. 
\end{equation}
As a result, we come to the final expression for the inverse localization
length, 
\begin{equation}
\label{locfin}l^{-1}={\frac{{\epsilon _0^2 \, \varphi(\mu) }} {8\,{\sin
{}^2\mu }}\,} ; \,\,\,\,\varphi(\mu )=1+2\sum\limits_{k=1}^\infty \xi
(k)\,\cos \,(2\mu \,k) \,\, . 
\end{equation}
Here, the function $\varphi \,(\mu )$ is given by the Fourier series with
the coefficients $\xi (k)\,$ which are the correlators of the site potential 
$\epsilon _n$.
Note that Eq. (\ref{locfin})
fails close to the band center, $E= 0,$ and to the band edges, $E= \pm
\,2\,\,$, where even the standard Anderson model exhibits peculiarities
(see, for example, \cite{IRT98} and references therein).

Eq. (\ref{locfin}) establishes the relation between the localization
length and correlations. It allows us to calculate the localization length
if statistical properties of the sequence $\epsilon _n$ are known. Note that
only binary correlators enter in Eq.~(\ref{locfin}). This property comes
from the Born approximation and not from restrictions for statistics of $%
\epsilon _n$ \cite{L89}. In particular, we do not assume that the statistics is
gaussian.

Formally, the same result has been derived by different methods 
in \cite {L89,GF88}.
For continuous models, the high-energy asymptotics of the Lyapunov
exponent is also determined by the correlation function of the
site-potential \cite{LGP88}.
To the best of our knowledge, the relation (\ref{locfin}) has not
been discussed yet concerning the problem of mobility edge.   
The advantage of the above approach based on the Hamiltonian 
representation, is that it  
allows a simple generalization to the Kronig-Penney model
with correlated disorder \cite{IK99} and to the models with off-diagonal
disorder. The approach also clarifies the physical meaning of the summation
in Eq.(\ref{locfin}). Namely, the localization appears as a result of
multiple scattering at different sites, see Eq.(\ref{mapbase}), and 
each scattering enters in the Born approximation, see Eq.(\ref{tetaN}).


To reveal the role of correlations, let us 
apply the general relation (\ref{locfin}) to some known models.
First, we consider the potential

\begin{equation}
\label{var}\epsilon _n=\epsilon _0\sqrt{2}\cos (2\pi \alpha n^\gamma ) 
\end{equation}
with $\alpha$ irrational. The correlation function $\xi (k)$ for this
sequence is given by the sum$\,$

\begin{equation}
\label{sum}\xi (k)=\lim _{N\rightarrow \infty }\,\frac 1N\sum_{n=1}^N\cos
(2\pi \gamma \alpha kn^{\gamma -1})\,\, . 
\end{equation}
For $0<\gamma <1$ all the correlators $\xi (k)$ are independent on $k$%
\thinspace \thinspace and equal $1$ , thus, giving $\varphi \,(\mu )= 0$ for 
$0 < \mu < \pi $. Therefore, for this case the result coincides with the
case of the constant potential $\epsilon _n=\epsilon _0$; in both cases the
inverse localization length vanishes. This is consistent with the
perturbative approach developed in \cite{GF88}.

For $\gamma \,$$=1$ Eq. (\ref{base}) is reduced to the Harper equation with
incommensurate potential. In this case the correlation function $\xi (k)$
oscillates, $\xi (k)=\cos (2\pi \alpha k)$, that also gives $\varphi \,(\mu
)=0$. Thus, in the Harper model with a weak potential all states are
extended \cite{S85,FGP92}. For $\gamma >1$ the correlators $\xi (k)$ vanish
and the localization length turns out to be the same as for the Anderson
model (for energy close to the band center, see \cite{T88}).

Let us consider now the random dimer model \cite{DWP90} which is specified
by the sequence $\epsilon _n$ having only two values $\epsilon _1$ and $%
\epsilon _2,$ each of them appears in pairs ($\epsilon _1\epsilon _1$ or $%
\epsilon _2\epsilon _2$) and each pair emerges with probability $1/2$ .
Fully transparent states are known to occur for $E=\epsilon _1$ and $%
E=\epsilon _2$ , see, for example, \cite{IKT95} and references therein.
Statistical properties of this model are characterized by the variance $%
\epsilon _0^2=\frac 12\left( \epsilon _1^2+\epsilon _2^2\right) $ and two
correlators, $\left\langle \epsilon _n\epsilon _{n-1}\right\rangle =\epsilon
_0^2\,\frac{3+2\lambda +3\lambda ^2}{4\left( 1+\lambda ^2\right) }$ and $%
\left\langle \epsilon _n\epsilon _{n-2}\right\rangle =\epsilon _0^2\frac{%
\left( 1+\lambda \right) ^2}{2\left( 1+\lambda ^2\right) }\,$ where $\lambda
=\epsilon _1/\epsilon _2$ . By substituting these correlators into Eq. (\ref
{locfin}), we obtain the inverse localization length for the dimer, 
\begin{equation}
\label{locdim}l^{-1}(E)=\frac{\epsilon _0^2\left( 1-\lambda \right) ^2}{%
8\left( 1+\lambda ^2\right) }\frac{E^2}{4-E^2}\,. 
\end{equation}

The generalization for $N$-mer when the values $\epsilon _1$ and $%
\epsilon _2$ appear in randomly distributed blocks of length $N$ , can be
readily done. For example, for the trimer we have: 
\begin{equation}
\label{loctrim}l^{-1}(E)=\frac{\epsilon _0^2\left( 1-\lambda \right) ^2}{%
12\left( 1+\lambda ^2\right) }\frac{\left( E^2-1\right) ^2}{4-E^2}\,. 
\end{equation}
Note that the above explicit expressions for dimer 
(\ref{locdim}) and trimer (\ref
{loctrim}) are given for the whole energy range; to the best of our
knowledge, this is a new result.

For the $N-$mer
constructed from the standard Anderson model by repeating $N$ times each
random value $\epsilon _n$, Eq.(\ref{locfin}) gives the same result obtained
in \cite{SVE94}. In particular, this result shows that
the correlations do not necessarily suppress the localization,
they can make it even stronger than in the Anderson model \cite{JK86}.

The localization length (\ref{locfin}) does not depend explicitly on the
site-potential but on the binary correlators $\xi (k)\,$of the potential.
Let us consider different types of correlations.
First, we take a sequence $\epsilon _n$ with exponential decay of
correlations, $\xi (k)=\exp (-\beta k)$ . This type of correlations is
known to occur in completely chaotic Hamiltonian models without stable
regions in the classical phase space. By substituting $\xi (k)$ into Eq. (%
\ref{locfin}), one gets the following expression,

\begin{equation}
\label{locexp}l^{-1}={\frac{\epsilon {_0^2}}{8{\sin {}^2\mu }}\,}{\frac{{%
\sinh \beta }}{{\cosh \beta -\cos (2\mu )}}}\,, 
\end{equation}
which establishes a link between the correlation radius (parameter $\beta
^{-1}$ ) of the site-potential, and the localization length of eigenstates.
In the limiting case $\beta \to \infty $, we get the result for uncorrelated
sequence (Anderson model) and for $\beta \to 0$ the result for the constant
potential is recovered.

Concerning the power decay of correlations, one should note that such simple
functions as $\xi (k)=1/k\,$or $1/k^2$ can not serve as the correlation
functions of any random sequence. Indeed, these functions lead to negative
values of the Lyapunov exponent and do not satisfy the condition required
for any correlation function \cite{KK68}. Specifically, the matrix $\xi
_{kk^{\prime }}\equiv \xi (k-k^{\prime })$ must be {\it positive-semidefinite},
the property which is not obvious {\it a priori} for arbitrary {\it positive}
function $\xi (k)$.

An important application of Eq. (\ref{locfin}) is related to the question
about existence of the mobility edge in 1D systems. So far the mobility edge
has been predicted for incommensurate potentials Eq.(\ref{var}) with $\gamma
<1$ and finite amplitude $\epsilon $$_0$ \cite{SHX88,FGP93}, and for
Kronig-Penney model with constant electric field \cite{DSS84}.

The relation (\ref{locfin}) 
allows us to construct site potentials with mobility edges. 
Rewritten in the form 
\begin{equation}
\label{ksi}\xi (k)=\frac 2\pi \int\limits_0^{\pi /2}\varphi \,(\mu )\,\cos
(2k\mu )\,d\mu 
\end{equation}
this relation gives a solution of the ``inverse problem'', namely, it shows
how to calculate the correlation function $\xi (k)$ of the random potential
if the normalized Lyapunov exponent $\Lambda _0\,(E)= \frac{2}{3}
\varphi \,(\arccos
(E/2))$ is known. In other words, for any dependence $\Lambda _0$ there
exist a set of sequences $\epsilon _n$ with correlation function
$\xi (k)$.

To reconstruct the sequence via the correlation function, we have used the
following algorithm \cite{G98}, 
\begin{equation}
\label{alg}\epsilon _n=A\sum\limits_{k=-\infty }^{+\infty }\xi (k)Z_{n+k}, 
\end{equation}
where $A$ is the normalization constant and $Z_s$ are random numbers from
the interval $[0,1]$. As an example, we consider two sequences $\epsilon _n$
which exhibit the mobility edges at $E=\pm 1$. The first one was obtained
from Eq.(\ref{alg}) by substituting $\xi (k)=\frac 3{2 \pi k}\sin 
(\frac{2\pi k%
}3)$ and $A=1$. This corresponds to the step-function dependence
of the Lyapunov exponent, $\Lambda _0=0$ for $|E|<1$ and $\Lambda _0=1$ for $%
1<|E|<2$. Circles in Fig. 1 show the dependence $\Lambda _0(E)$ calculated
numerically for the reconstructed sequence $\epsilon _n$ using Eq.~(\ref
{locdef}). The step-function is reproduced quite well and oscillations at $%
|E|>1$ are mainly due to the finite length of the sequence, $N=10^6$.

\begin{figure}
\vspace{-0.3cm}
\hspace {-1.2cm}
\epsfxsize 8cm
\epsfbox{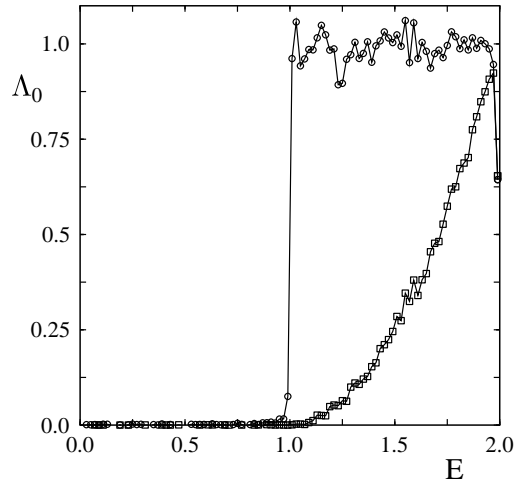}
\vspace{-0.5cm}
\narrowtext
\caption{Rescaled Lyapunov exponent $\Lambda_0(E)$
for two site-potentials $\epsilon_n$ constructed by Eq. 
(\ref{alg}) for $300$ correlators 
$\xi(k)=\xi(-k)$ and $n=1, ... , 10^6$. 
}
\label{fig1}
\end{figure}

Another sequence $\epsilon _n$ was generated in order to get a smooth
increase $\Lambda _0(E)$ for $|E|\geq 1$, see squares in Fig. 1. To do this,
we used the correlators $\xi (k)$ which are the Fourier components (\ref{ksi}%
) of the linear function $\Lambda _0(E)=1.53 (|E|-1)$ 
for $|E|\geq 1$. As one can
see, actual dependence $\Lambda _0(E)$ in Fig.~1 for $E>1$ clearly differs
from the linear one. This results from the fact that in general case the
algorithm (\ref{alg}) reconstructs the sequence $%
\epsilon _n$ with correlations $\xi (k)$ 
only asymptotically, for $k \gg 1$  \cite{G98}. Nevertheless, it allows us to
generate sequences with different values of critical exponents at the
mobility edge. These examples demonstrate the existence of mobility edges
for 1D site potentials with long-range correlations.

Another way to obtain explicitly the sequence $\epsilon _n$ with a mobility
edge is to use weakly chaotic non-linear maps. Let us take the
well-known {\it standard map} \cite{C79},
\begin{equation}
\label{stmap}
\begin{array}{c}
P_{n+1}=P_n+K\sin 
{2\pi X_n}\,\,\,\,\,\,(mod\,\,1) \\ X_{n+1}=X_n+P_{n+1}\,\,\,\,\,\,\,(mod\,%
\,1)
\end{array}
\end{equation}
For $K < 1$ there are trajectories close to separatrices of non-linear
resonance, with a very slow decay of correlations. For one of such
trajectories (see the inset in Fig. 2) we have constructed the sequence $%
\epsilon _n$ as follows, $\epsilon _n=2\epsilon _0\sin ({2\pi X_n}),$ and
calculated the Lyapunov exponent $\Lambda _0(E)=8\Lambda \sin^2 \mu /
\epsilon _0^2$ from Eq.(\ref{locdef}). The dependence given in Fig. 2 for $%
\epsilon _0=0.1$ clearly indicates the presence of two mobility edges. It is
important that the mobility edges are stable with respect to rather wide
variations of $\epsilon _0$. Independent computation of eigenenergies of Eq.
(\ref{base}) with the corresponding potential $\epsilon _n$ 
demonstrates that the
energy region shown in Fig. 2, indeed, belongs to the energy spectrum.

\begin{figure}
\vspace{-0.3cm}
\hspace {-1.2cm}
\epsfxsize 8cm
\epsfbox{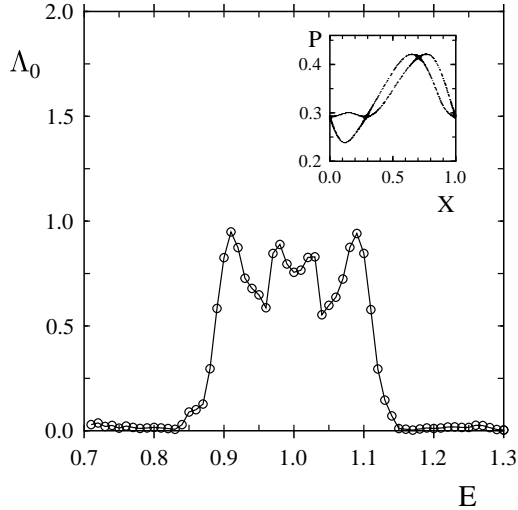}
\vspace{-0.5cm}
\narrowtext
\caption{Dependence $\Lambda_0(E)$ for the model
(\ref{map1}) with the site potential defined by Eq.(\ref{stmap}). The length
of the sequence $\epsilon_n$ is $N=10^6$. The corresponding trajectory
($X_n,\,P_n$) for $X_0=0, \, P_0=0.292 $ and $K=0.8$
is shown in the inset.
}
\label{fig2}
\end{figure}

It is interesting to note that inside a narrow chaotic region like that
shown in the inset of Fig.~2, the time dependence of trajectory
($X_n, \, P_n$) is given by regular rotation around the resonance of
period 3 with the following chaotic motion in the vicinities of the
crossed separatrices. This type of motion can be compared to 
the so-called intermittency \cite{MP80} which is well studied in 
1D maps. Therefore, one can expect that there is
a direct link between the correlations 
in the site potential and the intermittency
in the evolution of the corresponding dynamical model which generates 
this potential (see also \cite{GNS93}).

In conclusion, we have shown the existence of the mobility edge in 1D random
potentials. The relation (\ref{locfin}) allows us to determine 
binary correlation function of the site potential which exhibits any desirable
dependence of the localization length on energy. Specific problem we address 
in this Letter, is how to reconstruct the potential itself from
the binary correlations. We suggest an algorithm to calculate it numerically,
and
present the data for few specific potentials, with two mobility edges
inside the allowed energy zone. Potentials exhibiting mobility edges 
can be also constructed by making use of 
weakly chaotic trajectories of non-linear maps, this is demonstrated using 
the standard map. Although analytical results are derived in quadratic
approximation with random potential,
our numerical data clearly show that the mobility edges do not disappear
for relatively strong potentials. This fact opens a possibility 
for customer design  
of superlattices with anomalous electronic transport,
in particular, devices with selective transparancy.

Very recently, the so-called self-similar potentials with specific 
long-range correlations have been
studied \cite{ML98}, for which numerical data cleraly indicate the
exictence of mobility edges close to the center of the enery band.

We thank J. Bellissard, S. Gredeskul, I. Guarneri, M. Hilke, I. Satija
and V. Sokolov 
for useful discussions. This work was supported by CONACyT (Mexico) Grants
No. 26163-E and No.28626-E.

\end{multicols}

\end{document}